\documentclass[a4paper]{article}

\usepackage{INTERSPEECH2021}
\usepackage{multirow,hyperref}

\DeclareMathOperator*{\KL}{KL}

\DeclareMathOperator*{\diag}{diag}

\DeclareMathOperator*{\argmax}{argmax}

\title{Low-latency real-time non-parallel voice conversion based on cyclic variational autoencoder and
multiband WaveRNN with data-driven linear prediction}
\name{Patrick Lumban Tobing$^1$, Tomoki Toda$^1$}
\address{
  $^1$Nagoya University, Japan}
\email{patrick.lumbantobing@g.sp.m.is.nagoya-u.ac.jp, tomoki@icts.nagoya-u.ac.jp}

\begin{document}

\maketitle
\begin{abstract}
  This paper presents a low-latency real-time (LLRT) non-parallel voice conversion (VC) framework
  based on cyclic variational autoencoder (CycleVAE) and multiband WaveRNN with data-driven linear prediction (MWDLP).
  CycleVAE is a robust non-parallel multispeaker spectral model, which utilizes a speaker-independent latent space
  and a speaker-dependent code to generate reconstructed/converted spectral features given the spectral
  features of an input speaker. On the other hand, MWDLP is an efficient and a high-quality neural vocoder
  that can handle multispeaker data and generate speech waveform for LLRT applications with CPU. To accommodate
  LLRT constraint with CPU, we propose a novel CycleVAE framework that utilizes mel-spectrogram as spectral
  features and is built with a sparse network architecture. Further, to improve the modeling performance, we
  also propose a novel fine-tuning procedure that refines the frame-rate CycleVAE network by utilizing the
  waveform loss from the MWDLP network. The experimental results demonstrate that the proposed framework
  achieves high-performance VC, while allowing for LLRT usage with a single-core of $2.1$--$2.7$~GHz CPU on a
  real-time factor of $0.87$--$0.95$, including input/output, feature extraction, on a frame shift of
  $10$~ms, a window length of $27.5$~ms, and $2$ lookup frames.
\end{abstract}

\section{Introduction}

Voice conversion (VC) \cite{Childers85} is a technique for altering voice characteristics of a speech waveform
from an input speaker to that of a desired target speaker while preserving the lingustic contents of the speech.
Many real-world and/or research applications benefit from VC, such as for speech database augmentation
\cite{Abe90}, for recovery of impaired speech \cite{Tanaka13}, for expressive speech synthesis
\cite{Turk10}, for singing voice \cite{Kobayashi18}, for body-conducted speech processing
\cite{Toda12}, and for speaker verification \cite{Kinnunen12}. As the development of VC has
been growing rapidly \cite{Zhao20}, it is also wise to pursue not only for the highest
performance, but also for its feasibility on the constraints of real-world deployment/development, e.g.,
low-latency real-time (LLRT) \cite{Toda12b} constraint with low-computational machines in its deployment and
unavailability of parallel (paired) data between source and target speakers in its development.

To develop LLRT VC \cite{Toda12b}, the costs from input waveform analysis, conversion step, and output waveform
generation are taken into account to obtain the acceptable amount of total delay. On the waveform analysis,
several works use simple fast Fourier transform (FFT) \cite{Toda12b,Kobayashi20,Arakawa19,Saeki20}. On the
conversion module, where the spectral characteristics of speech waveform are usually modeled, a Gaussian mixture
model is employed in \cite{Toda12b}, a simple multi layer perceptron is employed in \cite{Arakawa19,Saeki20},
while convolutional neural network (CNN) and recurrent neural network (RNN) are employed in \cite{Kobayashi20}.
On the waveform generation, source-filter vocoder based on STRAIGHT \cite{Kawahara99} is used in
\cite{Toda12b,Kobayashi20}, while WORLD \cite{Morise16} is used in \cite{Arakawa19}, and direct waveform
filtering is utilized in \cite{Kobayashi18}. In all cases, parallel training data is required to develop the
conversion model, while the quality of the waveform generation module is still limited. In this paper, we work to
achieve flexible and high-quality LLRT VC, where it can be developed with non-parallel data and provide
high-quality waveform using also neural network for waveform generation, i.e., neural vocoder.

Neural vocoder could provide high-quality speech waveform in copy-synthesis \cite{Tamamori17},
in text-to-speech (TTS) \cite{Zhou20}, and in VC \cite{Zhao20} systems, albeit, high computational
cost impedes most of its use on LLRT applications. Essentially, neural vocoder architectures can be categorized
into autoregressive (AR) \cite{Kalchbrenner18,Valin19} and non-autoregressive (non-AR)
\cite{Prenger19,Yamamoto20} models, on which the former depends on the previously generated
waveform samples. In practice, AR models based on RNN (WaveRNN) \cite{Kalchbrenner18,Valin19} can be developed
with less layers than non-AR ones, which are built with multiple layers (deep) of CNN. In LLRT applications,
where waveform synthesis is sequentially performed depending on the availability of input stream, it is more
difficult for the deeper non-AR models to achieve this constraint while still preserving high-quality
waveform. In this work, to reliably achieve LLRT VC, we utilize a high-quality AR model called
multiband WaveRNN with data-driven linear prediction (MWDLP) \cite{Tobing21}, which has been proven to be capable
of producing high-fidelity waveform in the most adverse conditions including on LLRT constraint.

\begin{figure*}[!t]
  \centering
  \includegraphics[width=0.95\textwidth]{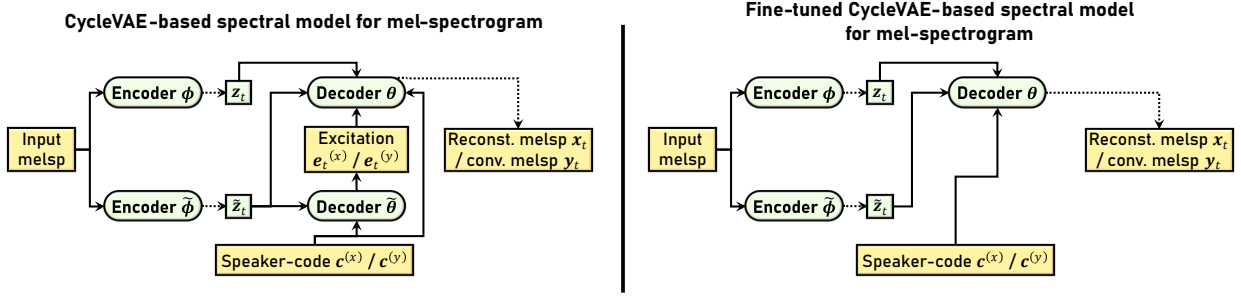}
  \vspace*{-2.5mm}
  \caption{Diagram of proposed CycleVAE model for mel-spectrogram (melsp) spectral features (left) with its
  fine-tuned architecture (right), where the second decoder $\tilde{\vec{\theta}}$ (excitation) is discarded,
  while keeping the related second encoder $\tilde{\vec{\phi}}$; Dotted lines denote sampling; Latent features
  are sampled from estimated posteriors; Reconstructed (reconst.) / converted (conv.) mel-spectrogram is sampled
  with estimated Gaussian parameters; Paths for speaker classifier (variational posterior of speaker-code)
  are omitted for brevity.}
  \vspace*{-4.5mm}
\label{fig:cycvae_base-ft}
\end{figure*}

On the other hand, to develop non-parallel VC, a shared space between speakers (speaker-independent)
can be utilized as a reference point on which the linguistic contents of speech are generated. For instance,
several works have employed the use of explicit text/phonetic space \cite{Huang20,Zhang20}.
An alternative way is to employ a linguistically unsupervised latent space that serves as a point of distribution
for the content generation, such as in variational autoencoder (VAE) \cite{Hsu16,Tobing19} or
generative adversarial network \cite{Kaneko19b}. The unsupervised approach has more flexibility
in terms of independency from linguistic features in its development, which could be of higher value in
situations where reliable transcriptions are difficult to be obtained. In this work, we focus on the use of a
robust model based on VAE called cyclic variational autoencoder (CycleVAE) \cite{Tobing20} that is capable of
handling non-parallel multispeaker data.

To achieve flexible and high-quality LLRT VC, we propose to combine CycleVAE-based spectral model
and MWDLP-based neural vocoder. First, we propose to modify the spectral features of CycleVAE to be
that of mel-spectrogram. Second, as in \cite{Kalchbrenner18,Valin19,Tobing21}, we propose to employ
sparsification for the CycleVAE network. Finally, to achieve high-performance VC, we propose a novel
fine-tuning for the CycleVAE model with the use of waveform domain loss from the MWDLP.
\section{MWDLP-based neural vocoder}

Let $\vec{s}=[s_1,\dotsc,s_{t_s},\dotsc,s_{T_s}]^{\top}$ be the sequence of speech waveform samples, where
$t_s$ and $T_s$ respectively denotes the time indices and the length of the waveform samples.
At band-level, the sequence of speech waveform samples is denoted as
$\vec{s}^{(m)}=[s^{(m)}_1,\dotsc,s^{(m)}_{\tau},\dotsc,s^{(m)}_{\mathcal{T}}]^{\top}$, where $m$ denotes the
$m$th band index, $\tau$ denotes the band-level time index, $\mathcal{T}=T_s/M$ denotes the length of the
band-level waveform samples, which is downsampled from $T_s$ by a factor of $M$ \cite{Nguyen94}, and the total number of
bands is denoted as $M$. At frame-level, the sequence of conditioning feature vectors is denoted as
$\vec{x}=[\vec{x}^{\top}_1,\dotsc,\vec{x}^{\top}_t,\dotsc,\vec{x}^{\top}_T]^{\top}$,
where $T$ denotes the length of the frame-level conditioning feature vector sequence, and at band-level,
the sequence of conditioning feature vectors is denoted as
$\tilde{\vec{x}}=[\tilde{\vec{x}}_1,\dotsc,\tilde{\vec{x}}_{\tau},\dotsc,\tilde{\vec{x}}_{\mathcal{T}}]$.

In MWDLP \cite{Tobing21}, the likelihood of the sequence of waveform samples $\vec{s}$ is defined
by the probability mass function (p.m.f.) of the discrete waveform samples as follows:
\begin{equation}
p(\vec{s}) =\!\! \prod_{m=1}^{M}\!\prod_{\tau=1}^{\mathcal{T}}
    p(s^{(m)}_{\tau}|\vec{s}^{(M)}_{1:\tau-1},\tilde{\vec{x}}_{\tau})
    =\!\! \prod_{m=1}^{M}\!\prod_{\tau=1}^{\mathcal{T}}
        \vec{p}^{(m)^{\top}}_{\tau}\!\!\!\vec{v}^{(m)}_{\tau},\!\!\!
\label{eq:mwdlp_like}
\end{equation}
where $\vec{s}^{(M)}_{1:\tau-1}$ denotes the past samples of all band-levels waveform,
$\vec{p}^{(m)}_{\tau}\!\!=\![p^{(m)}_{\tau}[1],\dotsc,p^{(m)}_{\tau}[b],\dotsc,p^{(m)}_{\tau}[B]]^{\top}\!\!\!$
denotes the probability vector, the number of sample bins is denoted as $B$, and $\vec{v}^{(m)}_{\tau}$
denotes a one-hot vector. Of the probability vector $\vec{p}^{(m)}_{\tau}$, the probability of each sample
bin $p^{(m)}_{\tau}[b]$ is given by
\begin{equation}
    p^{(m)}_{\tau}[b] = \frac{\exp(\hat{o}^{(m)}_{\tau}[b])}
            {\sum_{j=1}^{B}\exp(\hat{o}^{(_m)}_{\tau}[j])},
\label{eq:mwdlp_prob}
\end{equation}
where $\exp(\cdot)$ denotes the exponential function, $\hat{o}^{(m)}_{\tau}[b]$
is the unnormalized probability (logit) of the $b$th sample bin for the $m$th band,
and the vector of logits is denoted as  $\hat{\vec{o}}^{(m)}_{\tau}
=[\hat{o}^{(m)}_{\tau}[1],\dotsc,\hat{o}^{(m)}_{\tau}[b],\dotsc,\hat{o}^{(m)}_{\tau}[B]]^{\top}$.

The linear prediction (LP) \cite{Atal71} is performed in the logit space of the discrete waveform
samples as follows:
\begin{equation}
\hat{\vec{o}}^{(m)}_{\tau} = \sum_{k=1}^{K}a^{(m)}_{\tau}[k]\vec{r}^{(m)}_{\tau-k} + \vec{o}^{(m)}_{\tau},
\label{eq:mwdlp_lp}
\end{equation}
where the residual logit vector is denoted as $\vec{o}^{(m)}_{\tau}$, the $k$th data-driven LP coefficient of
the $m$th band is denoted as $a^{(m)}_{\tau}[k]$, $k$ denotes the index of LP coefficient, and
the total number of coefficients is denoted as $K$. $\{\vec{r}^{(m)}_{\tau-1},\dotsc,\vec{r}^{(m)}_{\tau-K}\}$
are the trainable logit basis vectors corresponding to past $K$ discrete samples. In Eq.~\ref{eq:mwdlp_lp},
the network outputs are $a^{(m)}_{\tau}[k]$ and $\vec{o}^{(m)}_{\tau}$.

\section{Proposed LLRT VC based on CycleVAE spectral model and MWDLP}

\subsection{CycleVAE model with mel-spectrogram features}
\label{ssec:cycvae_mel}

To realize LLRT VC, in this work, we propose to use mel-spectrogram as the spectral features for CycleVAE
model, where we extend the CycleVAE \cite{Tobing19,Tobing20} to incorporate estimation of intermediate excitation
features, e.g., fundamental frequency (F0). Diagram of the proposed model is illustrated in the left side
of Fig.~\ref{fig:cycvae_base-ft}.

Let $\vec{x}_t=[x_1[1],\dotsc,x_t[d],\dotsc,x_t[D]]^{\top}$ and
$\vec{y}_t=[y_1[1],\dotsc,y_t[d],\dotsc,y_t[D]]^{\top}$ be the $D$-dimensional spectral feature vectors
of an input speaker $x$ and that of a converted speaker $y$ at time $t$, respectively. The likelihood
function of the input spectral feature vector $\vec{x}_t$ is defined as follows:
\begin{align}
p_{\vec{\theta},\tilde{\vec{\theta}}}(\vec{x}_t,\vec{e}^{(x)}_t|\vec{c}^{(x)}_t) = \!\!\int\!\!\!\!\int\!
    &p_{\vec{\theta}}(\vec{x}_t|\vec{z}_t,\tilde{\vec{z}}_t,\vec{c}^{(x)}_t\!\!,\vec{e}^{(x)}_t)
        p_{\tilde{\vec{\theta}}}(\vec{e}^{(x)}_t|\tilde{\vec{z}}_t,\vec{c}^{(x)}_t) \nonumber \\
        &p_{\vec{\theta}}(\vec{z}_t)p_{\tilde{\vec{\theta}}}(\tilde{\vec{z}}_t)
            \mbox{d}\tilde{\vec{z}}_t\mbox{d}\vec{z}_t,\!\!\!
\label{eq:vae_mel_like}
\end{align}
where $\{\vec{z}_t,\tilde{\vec{z}}_t\}$ denotes the latent feature vectors, $\vec{c}^{(x)}_t$ denotes a
speaker-code vector of the input speaker $x$, and $\vec{e}^{(x)}_t$ denotes the excitation features. In VAE
\cite{Kingma13}, posterior form of latent features
$p_{\vec{\theta},\tilde{\vec{\theta}}}(\vec{z}_t,\tilde{\vec{z}}_t|\vec{x}_t)
    =\frac{p_{\vec{\theta},\tilde{\vec{\theta}}}(\vec{x}_t,\vec{z}_t,\tilde{\vec{z}}_t)}
        {p_{\vec{\theta},\tilde{\vec{\theta}}}(\vec{x}_t)}$
is utilized to handle the likelihood of Eq.~$\eqref{eq:vae_mel_like}$ with Gibbs' inequality as follows:
\begin{equation}
\log p_{\vec{\theta},\tilde{\vec{\theta}}}(\vec{x}_t,\vec{e}^{(x)}_t|\vec{c}^{(x)}_t)
    \geq \mathcal{L}(\vec{\Psi};\vec{x}_t,\vec{c}^{(x)}_t,\vec{e}^{(x)}_t),
\label{eq:vae_mel_form}
\end{equation}
where $\vec{\Psi}=\{\vec{\theta},\tilde{\vec{\theta}},\vec{\phi},\tilde{\vec{\phi}}\}$
and the variational/evidence lower bound (ELBO)
$\mathcal{L}(\vec{\Psi};\vec{x}_t,\vec{c}^{(x)}_t,\vec{e}^{(x)}_t)$ is given by
\begin{align}
&\mathbb{E}_{q_{\vec{\phi}\!,\tilde{\vec{\phi}}}(\vec{z}_t,\tilde{\vec{z}}_t|\vec{x}_t)}
    [\log p_{\vec{\theta}}(\vec{x}_t|\vec{z}_t,\!\tilde{\vec{z}_t},\!\vec{c}^{(x)}\!\!,\!\vec{e}^{(x)}_t)]
\!-\! \KL(q_{\vec{\phi}\!}(\vec{z}_t|\vec{x}_t)||p_{\vec{\theta}}(\vec{z}_t)) \nonumber \\
&+\! \mathbb{E}_{q_{\tilde{\vec{\phi}}}\!(\tilde{\vec{z}}_t\!|\vec{x}_t\!)}\!
    [\log p_{\tilde{\vec{\theta}}}\!(\vec{e}^{(x)}_t|\tilde{\vec{z}}_t,\!\vec{c}^{(x)}\!)]
    \!\!-\!\! \KL(q_{\tilde{\vec{\phi}}}\!(\tilde{\vec{z}}_t|\vec{x}_t)||p_{\tilde{\vec{\theta}}}(\tilde{\vec{z}}_t)\!),\!\!\!\!
\label{eq:vae_mel_elbo}
\end{align}
and $\vec{c}^{(x)}$ denotes a time-invariant speaker-code of the input speaker $x$. The sets of
encoder and decoder parameters are respectively denoted as $\{\vec{\phi},\tilde{\vec{\phi}}\}$
and $\{\vec{\theta},\tilde{\vec{\theta}}\}$. The prior distributions of latent features are denoted
as $p_{\vec{\theta}}(\vec{z}_t)$ and $p_{\tilde{\vec{\theta}}}(\tilde{\vec{z}}_t)$. The variational posteriors
are denoted as $q_{\vec{\phi}}(\vec{z}_t|\vec{x}_t)$ and $q_{\tilde{\vec{\phi}}}(\tilde{\vec{z}}_t|\vec{x}_t)$.
In addition, to improve the latent disentanglement performance, we also utilize variational posterior
$q_{\vec{\phi},\tilde{\vec{\phi}}}(\vec{c}^{(x)}_t|\vec{x}_t)$.

From Eq.~\eqref{eq:vae_mel_elbo}, the conditional probability density function (p.d.f.) of the input
spectral features $\vec{x}_t$, as well as of the converted spectral features $\vec{y}_t$, are given by
\begin{align}
p_{\vec{\theta}}(\vec{x}_t|\vec{z}_t,\tilde{\vec{z}}_t,\vec{c}^{(x)},\vec{e}^{(x)}_t) &= 
    \mathcal{N}(\vec{x}_t;\vec{\mu}^{(x)}_t,\vec{\Sigma}^{(x)}_t),
\label{eq:melsp_pdf_rec} \\
p_{\vec{\theta}}(\vec{y}_t|\vec{z}_t,\tilde{\vec{z}}_t,\vec{c}^{(y)},\vec{e}^{(y)}_t) &= 
    \mathcal{N}(\vec{y}_t;\vec{\mu}^{(y)}_t,\vec{\Sigma}^{(y)}_t), 
\label{eq:melsp_pdf_cv}
\end{align}
where $\vec{c}^{(y)}$ denotes the time-invariant speaker-code of the converted speaker $y$,
$\vec{e}^{(y)}_t$ denotes the converted excitation features, e.g., linearly converted log-F0 \cite{Toda07},
and
\begin{equation}
\vec{z}_t \!=\! \vec{\mu}^{(z)}_t \!- \vec{\sigma}^{(z)}_t\! \odot \vec{\epsilon},\,
    \tilde{\vec{z}}_t \!\!=\! \vec{\mu}^{(\tilde{z})}_t \!\!-\! \vec{\sigma}^{(\tilde{z})}_t\! \odot \vec{\epsilon},
    \,\vec{\epsilon} \sim\! \mathcal{L}(\vec{0},\!\vec{1}),\!\!\!
\label{eq:lat_est}
\end{equation}
the Hadamard product is denoted as $\odot$,  $\mathcal{L}(\vec{0},\vec{1})$ denotes the standard Laplacian
distribution. The Gaussian distribution with a mean vector $\vec{\mu}$ and a covariance matrix $\vec{\Sigma}$ is
denoted as $\mathcal{N}(;\vec{\mu},\vec{\Sigma})$. The output of encoders $\{\vec{\phi},\tilde{\vec{\phi}}\}$
are denoted as $\{\vec{\mu}^{(z)}_t,\vec{\sigma}^{(z)}_t,\vec{\mu}^{(\tilde{z})}_t,\vec{\sigma}^{(\tilde{z})}_t\}$,
while the output of decoder $\vec{\theta}$ is denoted as $\{\vec{\mu}^{(x)}_t,\diag(\vec{\Sigma}^{(x)}_t)\}$ or
$\{\vec{\mu}^{(y)}_t,\diag(\vec{\Sigma}^{(y)}_t)\}$. To improve the conversion performance, we also utilize the
p.d.f. of converted excitation $p_{\tilde{\vec{\theta}}}(\vec{e}^{(y)}_t|\tilde{\vec{z}}_t,\vec{c}^{(y)})$ in a
similar manner as in Eq.~\eqref{eq:vae_mel_elbo} of the excitation of input speaker. The reconstructed/converted
mel-spectrogram is generated from sampling the Gaussian p.d.f. in Eq.\eqref{eq:melsp_pdf_rec} or
\eqref{eq:melsp_pdf_cv}, respectively. 

\begin{figure}[!t]
  \centering
  \includegraphics[width=\linewidth]{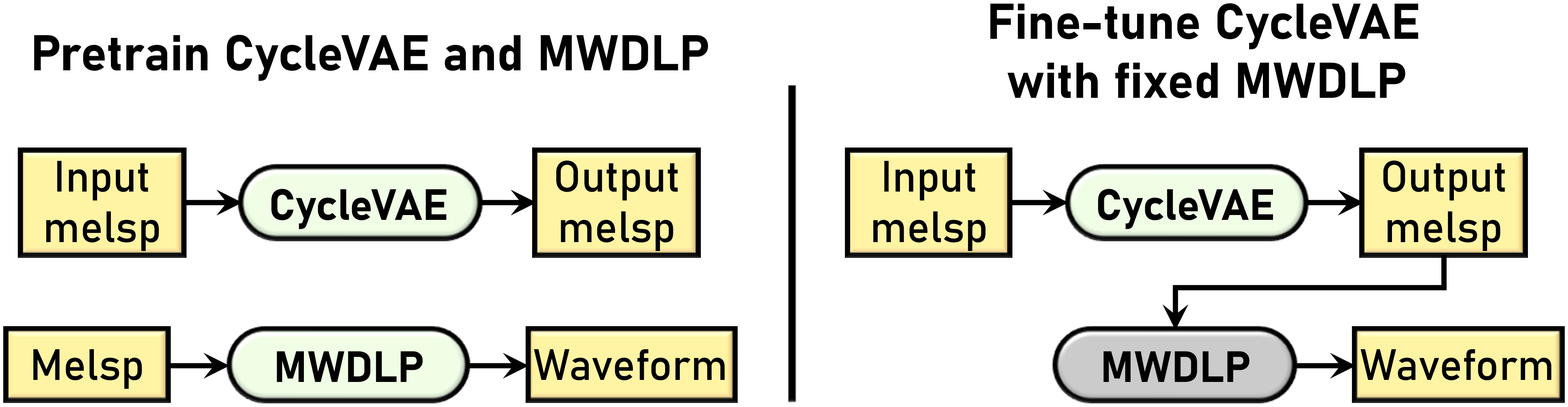}
  \vspace*{-3.5mm}
  \caption{Proposed model development steps: separately pretrain CycleVAE spectral model and MWDLP neural vocoder
  (left), then fine-tune CycleVAE modules with fixed MWDLP to utilize its waveform domain loss (right).}  
  \vspace*{-4.5mm}
\label{fig:cycvae_mwdlp}
\end{figure}

To provide network regularization with cycle-consistency, an auxiliary for the likelihood of
Eq.~\eqref{eq:vae_mel_like} is defined as follows:
\begin{align}
p_{\vec{\theta}\!,\tilde{\vec{\theta}}\!}(&\vec{x}_t,\vec{e}^{(x)}_t|\vec{e}^{(y)}_t,\vec{c}^{(x)}_t,\vec{c}^{(y)}_t) =
    \!\!\int\!\!\!\!\int\!\!\!\!\int\!
p_{\vec{\theta}}(\vec{x}_t|\vec{y}_t,\!\vec{z}_t,\!\tilde{\vec{z}}_t,\vec{e}^{(x)}_t\!\!,\vec{c}^{(x)}_t) \nonumber \\
&p_{\vec{\theta}}(\vec{y}_t|\vec{x}_t,\vec{z}_t,\tilde{\vec{z}}_t,\vec{e}^{(y)}_t,\vec{c}^{(y)}_t)
p_{\tilde{\vec{\theta}}}(\vec{e}^{(x)}_t|\tilde{\vec{z}}_t,\vec{c}^{(x)}_t)\mbox{d}\tilde{\vec{z}}_t\mbox{d}\vec{z}_t\mbox{d}\vec{y}_t,\!\!\!
\label{eq:aux_vae_cyc}
\end{align}
where by taking the expected values of the converted spectral $\vec{y}_t$ through sampling from
Eq.~\eqref{eq:melsp_pdf_cv}, Eq.~\eqref{eq:aux_vae_cyc} is rewritten as
\begin{align}
&p_{\vec{\theta}\!,\tilde{\vec{\theta}}\!}
    (\vec{x}_t,\!\vec{e}^{(x)}_t|\vec{e}^{(y)}_t\!\!,\vec{c}^{(x)}_t\!\!,\vec{c}^{(y)}_t\!) \!=\! \nonumber \\
    &\phantom{p_{\vec{\theta}\!,\tilde{\vec{\theta}}\!}(\vec{x}_t,}\!\!\int\!\!\!\!\!\int\!\!
\mathbb{E}_{p_{\vec{\theta}\!}(\vec{y}_t|\vec{x}_t,\vec{e}^{(y)}_t\!\!\!,\vec{c}^{(y)}_t)\!}
[p_{\vec{\theta}\!}
    (\vec{x}_t|\vec{y}_t,\!\vec{z}_t,\!\tilde{\vec{z}}_t,\vec{e}^{(x)}_t\!\!,\!\vec{c}^{(x)}_t\!)]
\nonumber \\
&\phantom{p_{\vec{\theta}\!,\tilde{\vec{\theta}}\!}(\vec{x}_t,\int\int}\;\:
p_{\tilde{\vec{\theta}}\!}(\vec{e}^{(x)}_t|\tilde{\vec{z}}_t,\vec{c}^{(x)}_t)
p_{\vec{\theta}}(\vec{z}_t)p_{\tilde{\vec{\theta}}\!}(\tilde{\vec{z}}_t)
\mbox{d}\tilde{\vec{z}}_t\mbox{d}\vec{z}_t.\!\!\!
\label{eq:aux_vae_cyc_exp}
\end{align}
Therefore, as in Eq.~\eqref{eq:vae_mel_form}, we approximate the true posterior
$p_{\vec{\theta},\tilde{\vec{\theta}}}
    (\vec{z}_t,\tilde{\vec{z}}_t|\vec{x}_t,\vec{y}_t)$ through the following form
\begin{equation}
\log p_{\vec{\theta}\!,\tilde{\vec{\theta}}\!}
    (\vec{x}_t,\!\vec{e}^{(x)}_t\!,\!\vec{e}^{(y)}_t|\vec{c}^{(x)}_t\!\!,\!\vec{c}^{(y)}_t\!) \!\geq\!
\mathcal{L}(\vec{\Psi};\!
\vec{x}_t,\!\vec{y}_t,\!\vec{e}^{(x)}_t\!\!,\!\vec{e}^{(y)}_t\!\!,\!\vec{c}^{(x)}_t\!\!,\!\vec{c}^{(y)}_t)\!\!\!
\label{eq:cycvae_mel_form}
\end{equation}
where the ELBO
$\mathcal{L}(\vec{\Psi};\!
\vec{x}_t,\!\vec{y}_t,\!\vec{e}^{(x)}_t\!\!,\!\vec{e}^{(y)}_t\!\!,\!\vec{c}^{(x)}_t\!\!,\!\vec{c}^{(y)}_t)$
is given by
\begin{align}
&\mathbb{E}_{p_{\vec{\theta}}(\vec{y}_t|\vec{x}_t,\vec{e}^{(y)}_t,\vec{c}^{(y)}_t)}
\big[\mathbb{E}_{q_{\vec{\phi}\!,\tilde{\vec{\phi}}\!}(\vec{z}_t\!,\tilde{\vec{z}}_t|\vec{x}_t,\vec{y}_t)}
    [\log p_{\vec{\theta}\!}(\vec{x}_t|\vec{z}_t,\!\tilde{\vec{z}_t},\!\vec{c}^{(x)}\!\!\!,\vec{e}^{(x)}_t)]]
\nonumber \\
&- \KL(q_{\vec{\phi}}(\vec{z}_t|\vec{x}_t,\vec{y}_t)||p_{\vec{\theta}}(\vec{z}_t))
- \KL(q_{\tilde{\vec{\phi}}}
    (\tilde{\vec{z}}_t|\vec{x}_t,\vec{y}_t)||p_{\tilde{\vec{\theta}}}(\tilde{\vec{z}}_t)) \nonumber \\
&+ \mathbb{E}_{q_{\tilde{\vec{\phi}}}\!(\tilde{\vec{z}}_t|\vec{y}_t)}
    [\log p_{\tilde{\vec{\theta}}}(\vec{e}^{(x)}_t|\tilde{\vec{z}}_t,\vec{c}^{(x)})]\big].
\label{eq:cycvae_mel_elbo}
\end{align}
Hence, the optimization of network parameters
$\hat{\vec{\Psi}}=\{\hat{\vec{\theta}},\hat{\tilde{\vec{\theta}}},\hat{\vec{\phi}},\hat{\tilde{\vec{\phi}}}\}$
is performed with Eqs.~\eqref{eq:vae_mel_form} and \eqref{eq:cycvae_mel_form} as follows:
\begin{align}
\hat{\vec{\Psi}}
    = \argmax_{\vec{\theta},\tilde{\vec{\theta}},\vec{\phi},\tilde{\vec{\phi}}} \sum_{t=1}^{T}
   &\mathcal{L}(\vec{\Psi};\!
\vec{x}_t,\!\vec{y}_t,\!\vec{e}^{(x)}_t\!\!,\!\vec{e}^{(y)}_t\!\!,\!\vec{c}^{(x)}_t\!\!,\!\vec{c}^{(y)}_t)
\nonumber \\
&+\mathcal{L}(\vec{\Psi};\vec{x}_t,\vec{c}^{(x)}_t,\vec{e}^{(x)}_t)
\label{eq:obj_cycvae}
\end{align}

\begin{figure}[!t]
  \centering
  \includegraphics[width=\linewidth]{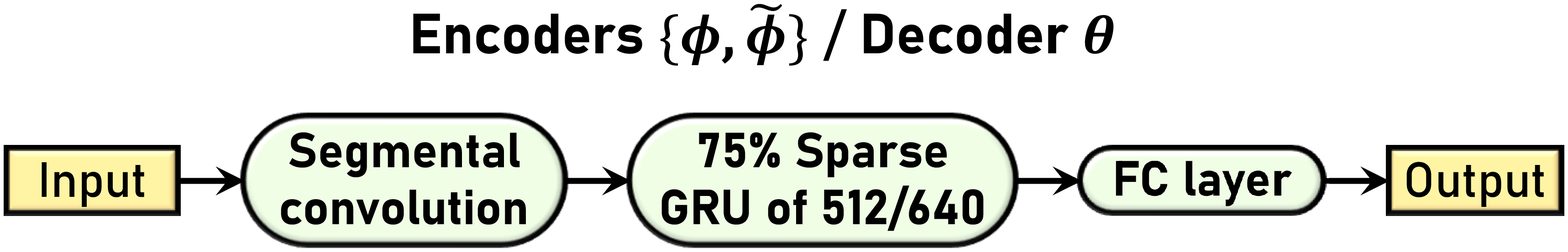}
  \vspace*{-3.5mm}
  \caption{Network structure of encoders $\{\vec{\phi},\tilde{\vec{\phi}}\}$ and decoder $\vec{\theta}$ with
  a base GRU size of $512$ and $640$, respectively, which are sparsified to $75\%$ density. Segmental convolution
  is made to take into account $p$ previous and $n$ succeeding frames, as in \cite{Tobing21},
  with $p=3,n=1$ and $p=4,n=0$ for encoders and decoders, respectively.}
  \vspace*{-4.5mm}
\label{fig:arch_enc-dec}
\end{figure}


\subsection{Fine-tuning with MWDLP-based waveform loss}
\label{ssec:cycvae_ft}

As illustrated on the right side of Fig.~\ref{fig:cycvae_base-ft} and Fig.~\eqref{fig:cycvae_mwdlp},
to perform fine-tuning with MWDLP loss, we discard the estimation of excitation, where the likelihood
in Eq.~\eqref{eq:vae_mel_like} is rewritten as follows:
\begin{equation}
p_{\vec{\theta}\!}(\vec{x}_t|\vec{c}^{(x)}_t) \!=\!\!
    \int\!\!\!\!\!\int\!\!p_{\vec{\theta}\!}(\vec{x}_t|\vec{z}_t,\!\tilde{\vec{z}}_t,\!\vec{c}^{(x)}_t)
    p_{\vec{\theta}\!}(\vec{z}_t)
        p_{\vec{\theta}\!}(\tilde{\vec{z}}_t)\mbox{d}\tilde{\vec{z}}_t\mbox{d}\vec{z}_t.\!\!\!\!
\label{eq:vae-ft_mel_like}
\end{equation}
As in Eqs.~\eqref{eq:vae_mel_form} and \eqref{eq:vae_mel_elbo}, the inequality form to approximate the true
posterior $p_{\vec{\theta}}(\vec{z}_t,\tilde{\vec{z}_t}|\vec{x}_t)$ is as follows:
\begin{equation}
\log p_{\vec{\theta}}(\vec{x}_t|\vec{c}^{(x)}_t) \geq \mathcal{L}(\vec{\Lambda};\vec{x}_t,\vec{c}^{(x)}_t)
\label{eq:vae-ft_mel_form}
\end{equation}
where $\vec{\Lambda}=\{\vec{\theta},\vec{\phi},\tilde{\vec{\phi}}\}$, and the ELBO
$\mathcal{L}(\vec{\Lambda};\vec{x}_t,\vec{c}^{(x)}_t)$ is given by
\begin{align}
\mathbb{E}_{q_{\vec{\phi}\!,\tilde{\vec{\phi}}\!}(\vec{z}_t\!,\tilde{\vec{z}}_t|\vec{x}_t)}
    [\log p_{\vec{\theta}\!}(\vec{x}_t|\vec{z}_t,&\tilde{\vec{z}_t},\vec{c}^{(x)})]
\!-\! \KL(q_{\vec{\phi}\!}(\vec{z}_t|\vec{x}_t||p_{\vec{\theta}\!}(\vec{z}_t)) \nonumber \\
&-\! \KL(q_{\tilde{\vec{\phi}}\!}(\tilde{\vec{z}}_t|\vec{x}_t)||p_{\vec{\theta}\!}(\tilde{\vec{z}}_t)).\!\!\!
\label{eq:vae-ft_mel_elbo}
\end{align}

Likewise, following Eq.~\eqref{eq:aux_vae_cyc_exp}, the auxiliary form of Eq.~\eqref{eq:vae-ft_mel_like},
to provide cycle-consistency, is defined as follows:
\begin{align}
p_{\vec{\theta}}
    (\vec{x}_t|\vec{c}^{(x)}_t\!\!,\vec{c}^{(y)}_t) =
    \!\!\int\!\!\!\!\!\int\!
&\mathbb{E}_{p_{\vec{\theta}}(\vec{y}_t|\vec{x}_t,\vec{c}^{(y)}_t)}
[p_{\vec{\theta}}
    (\vec{x}_t|\vec{y}_t,\!\vec{z}_t,\!\tilde{\vec{z}}_t,\!\vec{c}^{(x)}_t)]
\nonumber \\
&\phantom{\mathbb{E}_{p_{\vec{\theta}}(\vec{y}_t|\vec{x}_t}}
    p_{\vec{\theta}}(\vec{z}_t)p_{\vec{\theta}}(\tilde{\vec{z}}_t)
        \mbox{d}\tilde{\vec{z}}_t\mbox{d}\vec{z}_t.\!\!\!
\label{eq:aux_vae-ft_cyc_exp}
\end{align}
Following Eqs.~\eqref{eq:cycvae_mel_form} and \eqref{eq:cycvae_mel_elbo}, the inequality form
to approximate the true posterior $p_{\vec{\theta}}(\vec{z}_t,\tilde{\vec{z}_t}|\vec{x}_t,\vec{y}_t)$
is defined as
\begin{equation}
\log p_{\vec{\theta}}(\vec{x}_t|\vec{c}^{(x)}_t,\vec{c}^{(y)}_t) \geq
    \mathcal{L}(\vec{\Lambda};\vec{x}_t,\vec{y}_t,\vec{c}^{(x)}_t,\vec{c}^{(y)}_t)
\label{eq:cycvae-ft_mel_form}
\end{equation}
where the ELBO
$\mathcal{L}(\vec{\Lambda};\vec{x}_t,\vec{y}_t,\vec{c}^{(x)}_t,\vec{c}^{(y)}_t)$ is given by
\begin{align}
&\mathbb{E}_{p_{\vec{\theta}}(\vec{y}_t|\vec{x}_t,\vec{c}^{(y)}_t)}
\big[\mathbb{E}_{q_{\vec{\phi},\tilde{\vec{\phi}}}(\vec{z}_t,\tilde{\vec{z}}_t|\vec{x}_t,\vec{y}_t)}
    [\log p_{\vec{\theta}}(\vec{x}_t|\vec{z}_t,\tilde{\vec{z}_t},\vec{c}^{(x)})]]
\nonumber \\
&- \KL(q_{\vec{\phi}}(\vec{z}_t|\vec{x}_t,\vec{y}_t)||p_{\vec{\theta}}(\vec{z}_t))
- \KL(q_{\tilde{\vec{\phi}}}
    (\tilde{\vec{z}}_t|\vec{x}_t,\vec{y}_t)||p_{\vec{\theta}}(\tilde{\vec{z}}_t))\big].
\label{eq:cycvae-ft_mel_elbo}
\end{align}
Finally, the set of updated parameters
$\hat{\vec{\Lambda}}=\{\hat{\vec{\theta}},\hat{\vec{\phi}},\hat{\tilde{\vec{\phi}}}\}$ is obtained
by combining Eqs.~\eqref{eq:vae-ft_mel_form}, \eqref{eq:cycvae-ft_mel_form}, and
Eq.~\eqref{eq:mwdlp_like}, i.e., the likelihood of the waveform samples from MWDLP, as follows:
\begin{align}
\hat{\vec{\Lambda}}
    \!=\! \argmax_{\vec{\theta},\vec{\phi},\tilde{\vec{\phi}}} \!\sum_{t=1}^{T}
    &\mathcal{L}(\vec{\Lambda};\vec{x}_t,\vec{y}_t,\vec{c}^{(x)}_t,\vec{c}^{(y)}_t)
    \!+\!
    \mathcal{L}(\vec{\Lambda};\vec{x}_t,\vec{c}^{(x)}_t) \nonumber \\
&+\!\!\sum_{m=1}^{M}\!\sum_{\tau=1}^{\mathcal{T}}
    \log p(s^{(m)}_{\tau}|\vec{s}^{(M)}_{1:\tau-1},\tilde{\vec{x}}_{\tau}),\!\!\!
\label{eq:obj_cycvae_ft}
\end{align}
where the conditioning feature vector $\tilde{\vec{x}}_{\tau}$ is built from the sampled reconstructed
mel-spectrogram $\vec{x}_t$ of the input speaker $x$.

\subsection{Network architecture and sparsification}
\label{ssec:net_sparse}

The network architecture of the encoders and decoders of the proposed CycleVAE is illustrated in
Fig.\ref{fig:arch_enc-dec}. As in \cite{Tobing21}, a segmental convolution is utilized to take into
account $p$ preceding and $n$ succeeding frames. To realize LLRT VC, we use $p=3,n=1$ for encoders of CycleVAE,
$p=4,n=0$ for decoder of CycleVAE, and $p=5,n=1$ for the MWDLP neural vocoder, which yields a total of $2$
lookup frames.

In addition, a sparsification procedure for CycleVAE network is also performed, as in \cite{Valin19,Tobing21},
with $75\%$ target density for the gated recurrent unit (GRU) modules of encoders
$\{\vec{\phi},\tilde{\vec{\phi}}\}$ and decoder $\vec{\theta}$. The base hidden units size of GRU encoders is
$512$, while that of the decoder is $640$. The target density ratios for each reset, update, and new gates of the
GRU recurrent matrices are respectively $0.685,0.685,0.88$.

\begin{table}[!t]
  \scriptsize
  \caption{Results of accuracy (acc.) measurement on log-global-variance distance of mel-cepstrum (LGD),
  mel-cepstral distortion (MCD), unvoiced/voiced decision error (U/V), and root-mean-square-error of F0 between
  converted and target waveform on intra-lingual pairs.}
  \vspace{-2.5mm}
  \centering
  \begin{tabular}{c c c c c c}
    \toprule
    \multicolumn{1}{c}{\textbf{Intra-lingual acc.}}
    & \multicolumn{1}{c}{\textbf{LGD}}
        & \multicolumn{1}{c}{\textbf{MCD [dB]}}
            & \multicolumn{1}{c}{\textbf{U/V [\%]}}
                & \multicolumn{1}{c}{\textbf{F0 [Hz]}} \\
    \midrule
    \textbf{ASR+TTS} \cite{Huang20}                           & $0.29$~~~      & $6.91$~~~           & $16.20$~~~     & $\textbf{22.29}$~~~          \\
    \textbf{CycVAE+PWG} \cite{Tobing20}                      & $0.34$~~~      & $\textbf{6.67}$~~~           & $\textbf{14.36}$~~~     & $24.91$~~~               \\
    \textbf{NU T23} \cite{Huang20b}                      & $\textbf{0.28}$~~~      & $7.50$~~~           & $18.94$~~~     & $23.20$~~~               \\
    \midrule
    \textbf{LLRT CycVAE}              & $0.36$~~~      & $\textbf{7.41}$~~~           & $\textbf{15.35}$~~~     & $25.74$~~~          \\
    \textbf{LLRT CycVAE+FT}            & $\textbf{0.28}$~~~       & $7.51$~~~           & $17.27$~~~    & $\textbf{25.17}$~~~           \\
    \bottomrule
  \end{tabular}
  \vspace{-1.5mm}
\label{tab:obj_res_intra}
\end{table}

\begin{table}[!t]
  \scriptsize
  \caption{Results of accuracy (acc.) measurement on log-global-variance distance of mel-cepstrum (LGD),
  mel-cepstral distortion (MCD), unvoiced/voiced decision error (U/V), and root-mean-square-error of F0 between
  converted and target waveform on cross-lingual pairs.}
  \vspace{-2.5mm}
  \centering
  \begin{tabular}{c c c c c c}
    \toprule
    \multicolumn{1}{c}{\textbf{Cross-lingual acc.}}
    & \multicolumn{1}{c}{\textbf{LGD}}
        & \multicolumn{1}{c}{\textbf{MCD [dB]}}
            & \multicolumn{1}{c}{\textbf{U/V [\%]}}
                & \multicolumn{1}{c}{\textbf{F0 [Hz]}} \\
    \midrule
    \textbf{ASR+TTS} \cite{Huang20}                           & $0.39$~~~      & $8.78$~~~           & $14.84$~~~     & $\textbf{21.12}$~~~          \\
    \textbf{CycVAE+PWG} \cite{Tobing20}                      & $0.34$~~~      & $\textbf{7.56}$~~~           & $\textbf{13.86}$~~~     & $22.83$~~~               \\
    \textbf{NU T23} \cite{Huang20b}                      & $\textbf{0.24}$~~~      & $8.50$~~~           & $16.33$~~~     & $22.68$~~~               \\
    \midrule
    \textbf{LLRT CycVAE}              & $0.39$~~~      & $\textbf{8.22}$~~~           & $\textbf{15.25}$~~~     & $20.91$~~~          \\
    \textbf{LLRT CycVAE+FT}            & $\textbf{0.30}$~~~       & $8.44$~~~           & $15.81$~~~    & $\textbf{20.91}$~~~           \\
    \bottomrule
  \end{tabular}
  \vspace{-4.5mm}
\label{tab:obj_res_cross}
\end{table}

\section{Experimental evaluation}

\subsection{Experimental conditions}
\label{ssec:exp_cond}

We used the Voice Conversion Challenge (VCC) 2020 \cite{Zhao20} dataset, which consisted of $8$ English speakers,
$2$ German speakers, $2$ Finnish speakers, and $2$ Mandarin speakers, each uttered $70$ sentences in their
languages. For the training set, $60$ sentences were used, while the remaining $10$ sentences were for
the development set. Additional $25$ English utterances from each speaker were provided for evaluation. In the
evaluation, we utilized two baseline systems of VCC 2020: cascaded automatic speech recognition (ASR) with TTS
(ASR+TTS) \cite{Huang20} and CycleVAE with Parallel WaveGAN (CycVAE+PWG) \cite{Tobing20}, as well as Nagoya
University (NU) T23 system \cite{Huang20b}. $2$ English source, $2$ English target (intra-lingual), and
$2$ German target (cross-lingual) speakers were utilized in the evaluation.

As spectral features, we used $80$-dimensional mel-spectrogram, which was extracted frame-by-frame from
magnitude spectra. The number of FFT length in analysis was $2048$. $27.5$~ms Hanning window with $10$~ms frame
shift were used. The sampling rate was $24,000$~Hz. As the target intermediate excitation features used in
Section~\ref{ssec:cycvae_mel}, we used F0, aperiodicities, and their voicing decisions, which were
extracted from the speech waveform using WORLD \cite{Morise16}. The excitation $\vec{e}^{(y)}_t$
of converted speaker $y$ was set to linearly converted log-F0 \cite{Toda07}.

Other than the configuration of segmental convolution in Section~\ref{ssec:net_sparse}, the hyperparameters
of MWDLP neural vocoder was the same as in \cite{Tobing21} with the use of $K=8$ data-driven LP coefficients
and STFT loss. As well as for the CycleVAE-based spectral model, the encoders
$\{\vec{\phi},\tilde{\vec{\phi}}\}$ and the decoder $\vec{\theta}$ were set the same as in
\ref{ssec:net_sparse}. On the other hand, the excitation decoder $\tilde{\vec{\theta}}$ described
in Section~\ref{ssec:cycvae_mel} used the same structure as the other encoders/decoder, but utilizing a
dense GRU with $128$ hidden units. A classifier network with similar structure utilizing a GRU with $32$
hidden units was employed to handle the variational speaker posteriors
$q(\vec{c}^{(x)}_t|\vec{x}_t)$ and $q(\vec{c}^{(y)}_t|\vec{y}_t)$. Additionally, each of the encoders was
also set to estimate the speaker posteriors along with the latent posteriors.

The training procedure was as described in Sections~\ref{ssec:cycvae_mel} and \ref{ssec:cycvae_ft}, where
the standard Laplacian prior was replaced with the posterior of the pretrained CycleVAE. In addition, we
performed final fine-tuning of CycleVAE by fixing the encoders and updating only decoder $\vec{\theta}$
(LLRT CycVAE+FT). In all CycleVAE optimizations, the spectral loss included Gaussian p.d.f. term
and the loss of the sampled mel-spectrogram. Further, in the fine-tuning steps, we included loss from
full-resolution magnitude spectra, which was obtained using inverted mel-filterbank and the sampled
mel-spectrogram. The waveform domain loss included the set of loss in \cite{Tobing21} and the differences
of the output of all MWDLP layers when fed with original spectra and generated spectra (layer-wise loss).

We used a single-core of Intel Xeon Gold 6230 $2.1$~GHz, Intel Xeon Gold 6142 $2.6$~GHz, and
Intel i7-7500U $2.7$~GHz CPUs to measure the real-time factor (RTF), which respectively yield
$0.87$, $0.87$, and $0.95$ RTFs. The total delay is $23.75$~ms, which was the sum of the half
of the window length (1st frame) and one frame shift, i.e., $2$ lookup frames. The model development
software, real-time implementation, and audio samples have been made available at
{\scriptsize{\url{https://github.com/patrickltobing/cyclevae-vc-neuralvoco}}\normalsize.

\begin{table}[!t]
  \scriptsize
  \caption{Result on automatic speech recognition accuracy (ASR acc.) on intra- and cross-lingual conversions
  with word error rate (WER) and character error rate (CER) measurements.}
  \vspace{-2.5mm}
  \centering
  \begin{tabular}{c c c c c c}
    \toprule
    \multirow{2}{*}{\textbf{ASR acc.}}
    & \multicolumn{2}{c}{\textbf{Intra-lingual}}
    && \multicolumn{2}{c}{\textbf{Cross-lingual}} \\
     & \multicolumn{1}{c}{\textbf{WER}}
        & \multicolumn{1}{c}{\textbf{CER}}
       && \multicolumn{1}{c}{\textbf{WER}}
        & \multicolumn{1}{c}{\textbf{CER}} \\
    \midrule
    \textbf{Source}                           & $18.5$~~~      & $3.7$~~~           && -~~~     & -~~~          \\
    \textbf{Target}                           & $17.5$~~~      & $3.0$~~~           && $19.2$~~~     & $4.1$~~~          \\
    \midrule
    \textbf{ASR+TTS} \cite{Huang20}                           & $\textbf{25.1}$~~~      & $\textbf{7.5}$~~~           && $30.3$~~~     & $12.2$~~~          \\
    \textbf{CycVAE+PWG} \cite{Tobing20}                      & $28.2$~~~      & $9.6$~~~           && $29.6$~~~     & $10.3$~~~               \\
    \textbf{NU T23} \cite{Huang20b}                      & $37.3$~~~      & $14.9$~~~           && $\textbf{25.2}$~~~     & $\textbf{7.6}$~~~
                   \\
    \midrule                  
    \textbf{LLRT CycVAE}              & $33.8$~~~      & $13.6$~~~           && $34.0$~~~     & $12.4$~~~          \\
    \textbf{LLRT CycVAE+FT}            & $\textbf{25.2}$~~~       & $\textbf{7.9}$~~~          && $\textbf{26.1}$~~~    & $\textbf{7.9}$~~~           \\
    \bottomrule
  \end{tabular}
  \vspace{-1.5mm}
\label{tab:obj_res_asr}
\end{table}

\subsection{Objective evaluation}

In the objective evaluation, we measured the accuracies of the generated waveforms to the target ground truth
and the accuracies of automatic speech recognition (ASR) output. The former was measured with the use of
mel-cepstral distortion (MCD), root-mean-square error of F0, unvoiced/voiced decision error (U/V), and
log of global variance \cite{Toda07} distance of the mel-cepstrum (LGD). The latter was measured with word error
rate (WER) and character error rate (CER). $28$-dimensional mel-cepstral coefficients were extracted from WORLD
\cite{Morise16} spectral envelope to compute the MCD. For ASR, we used ESPnet's \cite{Watanabe18} latest
pretrained model on LibriSpeech \cite{Panayotov15} data.

The results on the accuracies of the generated waveforms are shown in Tables~\ref{tab:obj_res_intra} and
\ref{tab:obj_res_cross}, which correspond to the intra- and the cross-lingual conversion pairs, respectively.
It can be observed that the proposed LLRT system based on CycleVAE and MWDLP utilizing fine-tuning with
waveform domain loss (LLRT CycVAE+FT) achieves better LGD values (less oversmoothed) in intra- and
cross-lingual conversions than the proposed system without fine-tuning (LLRT CycleVAE) with values of $0.28$
and $0.36$, respectively, in intra-lingual, and $0.30$ and $0.39$, respectively, in cross-lingual. Furthermore,
it beats the LGD values of CycVAE+PWG that uses non-LLRT system, and beats NU T23 system in MCD, U/V, and F0 for
cross-lingual, which uses non-LLRT CycleVAE with WaveNet.

Lastly, the ASR result is shown in Table~\ref{tab:obj_res_asr}, which shows WERs and CERs for the intra- and the
cross-lingual conversions. It can be clearly observed that the proposed LLRT CycVAE+FT outperforms the proposed
system without fine-tuning LLRT CycVAE with WER and CER values of $26.1$ and $7.9$ in intra-lingual and of
$25.2$ and $7.9$ in cross-lingual. These values are also lower than the non-LLRT CycleVAE system of the
VCC 2020 baseline (CycVAE+PWG) and similar to that of the non-LLRT CycleVAE of NU T23 in cross-lingual
conversions.

\subsection{Subjective evaluation}

In the subjective evaluation, we conducted two listening tests, each to judge the naturalness of speech
waveform and the speaker similarity to a reference target speech. The former is conducted with a mean
opinion score (MOS) test using a 5-scaled score ranging from 1 (very bad) to 5 (very good). The latter is
conducted with a speaker similarity test as in \cite{Zhao20}, where "same" or "not-same" decision had
to be chosen along with "sure" or "not-sure" decision as a confidence measure. $10$ utterances from
the evaluation set was used. The number of participants on Amazon Mechanical Turk was $19$ and $13$,
respectively, for MOS and speaker similarity tests.

\begin{table}[!t]
  \scriptsize
  \caption{Result of mean opinion score (MOS) test on naturalness for intra- and cross-lingual
  conversions in same-gender (SGD) and cross-gender (XGD) pairs. $^{*}$ denotes systems with statistically
  significant different values ($\alpha < 0.05$) compared to LLRT CycleVAE+FT in each conversion categories.}
  \vspace{-2.5mm}
  \centering
  \begin{tabular}{c c c c c c}
    \toprule
    \multirow{2}{*}{\textbf{MOS}}
    & \multirow{2}{*}{\textbf{All}}
    & \multicolumn{2}{c}{\textbf{Intra-lingual}}
    & \multicolumn{2}{c}{\textbf{Cross-lingual}} \\
     & & \multicolumn{1}{c}{\textbf{SGD}}
        & \multicolumn{1}{c}{\textbf{XGD}}
       & \multicolumn{1}{c}{\textbf{SGD}}
        & \multicolumn{1}{c}{\textbf{XGD}} \\
    \midrule
    \textbf{Source}               & $4.68$            & -~~~      & -~~~           & -~~~     & -~~~  \\
    \textbf{Target}               & $4.69$            & -~~~      & -~~~           & -~~~     & -~~~  \\
    \midrule
    \textbf{ASR+TTS} \cite{Huang20}               & $4.01$            & $\textbf{4.32}^{*}$~~~      & $4.15^{*}$~~~           & $3.84$~~~     & $3.72$~~~          \\
    \textbf{CycVAE+PWG} \cite{Tobing20}          & $3.85^{*}$            & $3.85^{*}$~~~      & $3.75$~~~           & $3.94$~~~     & $3.87$~~~               \\
    \textbf{NU T23} \cite{Huang20b}            & $\textbf{4.23}^{*}$          & $4.30^{*}$~~~      & $\textbf{4.21}^{*}$~~~           & $\textbf{4.23}^{*}$~~~     & $\textbf{4.21}^{*}$~~~               \\
    \midrule
    \textbf{LLRT CycVAE}         & $3.33^{*}$     & $3.30^{*}$~~~      & $3.19^{*}$~~~           & $3.48^{*}$~~~     & $3.19^{*}$~~~          \\
    \textbf{LLRT CycVAE+FT}      & $\textbf{3.96}$      & $\textbf{3.99}$~~~       & $\textbf{3.85}$~~~          & $\textbf{4.02}$~~~    & $\textbf{3.96}$~~~           \\
    \bottomrule
  \end{tabular}
  \vspace{-1.5mm}
\label{tab:sub_res_mos}
\end{table}

The result of MOS test on naturalness is shown in Table~\ref{tab:sub_res_mos}. It can be observed
that the proposed LLRT VC system benefits from the fine-tuning approach (LLRT CycleVAE+FT),
yielding significantly higher naturalness in all categories than the LLRT CycleVAE system,
with values of $3.96$, $3.99$, $3.85$, $4.02$, and $3.96$ for all, intra-lingual same-gender (SGD),
intra-lingual cross-gender (XGD), cross-lingual SGD and cross-lingual XGD, respectively. On the
other hand, the result of speaker similarity test is shown in Table~\ref{tab:subj_res_sim}. The
tendency is also similar, where the proposed LLRT CycleVAE+FT system has better speaker accuracy
than the LLRT CycleVAE in all categories, while achieving similar accuracies to the non-LLRT
CycleVAE systems: CycVAE+PWG and NU T23 (cross-lingual).

\begin{table}[!t]
  \scriptsize
  \caption{Result of speaker similarity [\%] test for intra- and cross-lingual conversions in same-gender (SGD)
  and cross-gender (XGD) pairs. $^{*}$ denotes systems with statistically significant different values
  ($\alpha < 0.05$) compared to LLRT CycleVAE+FT in each conversion categories.}
  \vspace{-2.5mm}
  \centering
  \begin{tabular}{c c c c c c}
    \toprule
    \multirow{1}{*}{\textbf{Speaker}}
    & \multirow{2}{*}{\textbf{All}}
    & \multicolumn{2}{c}{\textbf{Intra-lingual}}
    & \multicolumn{2}{c}{\textbf{Cross-lingual}} \\
     \multirow{1}{*}{\textbf{similarity [\%]}} & & \multicolumn{1}{c}{\textbf{SGD}}
        & \multicolumn{1}{c}{\textbf{XGD}}
       & \multicolumn{1}{c}{\textbf{SGD}}
        & \multicolumn{1}{c}{\textbf{XGD}} \\
    \midrule
    \textbf{Source}               & $8.01$            & -~~~      & -~~~           & -~~~     & -~~~  \\
    \textbf{Target}               & $90.05$            & -~~~      & -~~~           & -~~~     & -~~~  \\
    \midrule
    \textbf{ASR+TTS} \cite{Huang20}               & $\textbf{89.43}^{*}$            & $91.80$~~~      & $87.10^{*}$~~~           & $\textbf{84.12}^{*}$~~~     & $\textbf{87.90}^{*}$~~~          \\
    \textbf{CycVAE+PWG} \cite{Tobing20}          & $78.63$            & $85.25$~~~      & $77.42$~~~           & $74.19$~~~     & $77.78$~~~               \\
    \textbf{NU T23} \cite{Huang20b}            & $80.24^{*}$          & $\textbf{93.50}^{*}$~~~      & $\textbf{89.60}^{*}$~~~           & $71.77$~~~     & $66.13^{*}$~~~               \\
    \midrule
    \textbf{LLRT CycVAE}         & $70.22$     & $76.99$~~~      & $70.49$~~~           & $67.20$~~~     & $66.13^{*}$~~~          \\
    \textbf{LLRT CycVAE+FT}      & $\textbf{77.55}$      & $\textbf{86.18}$~~~       & $\textbf{74.16}$~~~          & $\textbf{75.24}$~~~    & $\textbf{74.60}$~~~           \\
    \bottomrule
  \end{tabular}
  \vspace{-1.5mm}
\label{tab:subj_res_sim}
\end{table}

\section{Discussion}

The proposed method of fine-tuning the CycleVAE-based spectral model with MWDLP-based waveform
modeling significantly improves the converted speech waveform. From our investigation, the use of
mel-spectrogram sampling from Gaussian p.d.f. in Eqs.\eqref{eq:melsp_pdf_rec} and
\eqref{eq:melsp_pdf_cv} works very well with the waveform domain loss. In addition, we also found
that layer-wise loss from neural vocoder helps to provide more natural outcome. Our reasoning is
that the generated spectra will not be exactly the same as the natural spectra that corresponds to
the natural waveform, but we assume that there is a domain for generated spectra that could provide
quite reasonable approximation for generating the natural waveform by explicitly guiding through
all layers of the neural vocoder in addition of the waveform loss.

The largest average RTF factors for each module are as follows: $0.14$ for two encoders, $0.13$ for
decoder, $0.56$ for MWDLP, and $0.12$ for others including input/output, memory allocation, etc.
The total of these RTF values, i.e., $\sim\!\!9.5$~ms, should be lower than the length of the frame
shift, which is $10$~ms. However, in practical situation, a larger margin is required to avoid
glitching caused by outliers of RTF values that are larger than the frame shift. In future work, we
will investigate lower size of MWDLP and/or $8$-bit model quantization.

\section{Conclusions}

We have presented a novel low-latency real-time (LLRT) non-parallel voice conversion (VC) framework
based on cyclic variational autoencoder (CycleVAE) and multiband WaveRNN with data-driven linear
prediction (MWDLP). The proposed system utilizes mel-spectrogram features as the spectral
parameters of the speech waveform, which are used in the CycleVAE-based spectral model and the
MWDLP neural vocoder. To realize LLRT VC, CycleVAE modules undergo a sparsification procedure with
respect to their recurrent matrices. In addition, we propose to use waveform domain loss from a
fixed pretrained MWDLP to fine-tune the CycleVAE modules. The experimental resuts have demonstrated
that the proposed system is capable of achieving high-performance VC, while allowing its usage for
LLRT applications with $0.87$--$0.95$ real-time factor using a single-core of $2.1$--$2.7$~GHz CPU
on $27.5$~ms window length, $10$~ms frame shift, and $2$ lookup frames.

\section{Acknowledgements}

This work was partly supported by JSPS KAKENHI Grant Number 17H06101 and JST, CREST Grant Number
JPMJCR19A3.


\bibliographystyle{IEEEtran}

\bibliography{rtvc_cycvae_mwdlp}


\end{document}